%% file: main.tex
\documentclass[10pt,a4paper]{article}

\usepackage{graphicx}%,color}
\usepackage{hyperref} 
\usepackage{amsmath,amsfonts}%
\usepackage{tabularx}
\usepackage{enumerate}
\usepackage{algorithm}
\usepackage{algpseudocode}
\usepackage{xspace}
\usepackage{authblk}
\usepackage{multirow}
\newtheorem{defin}{Definition}
\newtheorem{prop}{Proposition}

\usepackage{longtable}
\usepackage{graphicx}
\usepackage[colorinlistoftodos]{todonotes}
\usepackage{subfigure} % subfiguras
\usepackage[utf8]{inputenc}

\def\E{\mathcal{E}}
\def\B{\mathcal{B}}
\def\V{\mathcal{V}}
\def\H{\mathcal{H}}
\def\D{\mathcal{D}}
\def\I{\mathbb {I}}
\def\R{\mathbb{R}}
\def\A{\mathcal{A}}
\def\II{\mathcal{I}}
\def\L{\mathcal{L}}
\def\S{\mathcal{S}}

\graphicspath{{Figures/}} % Directory in which figures are stored

\def\S{\mathcal{S}}
\def\D{\mathcal{D}}
\def\H{\mathcal{H}}

\def\M{\mathcal{M}}

\providecommand{\keywords}[1]
{
  \small	
  \textbf{\textit{Keywords---}} #1
}

\definecolor{lime}{HTML}{A6CE39}
\DeclareRobustCommand{\orcidicon}{%
	\begin{tikzpicture}
	\draw[lime, fill=lime] (0,0) 
	circle [radius=0.16] 
	node[white] {{\fontfamily{qag}\selectfont \tiny ID}};
	\draw[white, fill=white] (-0.0625,0.095) 
	circle [radius=0.007];
	\end{tikzpicture}
	\hspace{-2mm}
}

\foreach \x in {A, ..., Z}{%
	\expandafter\xdef\csname orcid\x\endcsname{\noexpand\href{https://orcid.org/\csname orcidauthor\x\endcsname}{\noexpand\orcidicon}}
}

\title{ Data analysis  and the metric evolution of hypergraphs }

\author[1,3]{Dalma Bilbao }
\author[1,3]{Hugo Aimar }
\author[1,2,3,*] {Diego M. Mateos\orcidA{}}

\affil[1]{\normalsize Consejo Nacional de Investigaciones Cient\'ificas y T\'ecnicas (CONICET), Argentina.}
\affil[2]{\normalsize Facultad de Ciencia y Tecnolog\'{\i}a. Universidad Aut\'{o}noma de Entre R\'{\i}os (UADER). Oro Verde, Entre R\'{\i}os, Argentina.}
\affil[3]{\normalsize Instituto de Matem\'{a}tica Aplicada del Litoral (IMAL-CONICET-UNL), CCT CONICET, Santa F\'e, Argentina.}

\affil[*]{Corresponding author: Diego M. Mateos, mateosdiego@gmail.com}
\begin{document}
\date{}
\maketitle

\begin{abstract}

In this paper we aim to use different metrics in the Euclidean space and Sobolev type metrics in function spaces in order to produce reliable parameters for the  differentiation of point distributions and dynamical systems. The main tool is the analysis of the geometrical evolution of the hypergraphs generated by the growth of the radial parameters for a choice of an appropriate metric in the space containing the data points. Once this geometric dynamics is obtained we use Lebesque and Sobolev type norms in order to compare the basic geometric signals obtained.  

\end{abstract}


%==================================================
%                   Introduction
%==================================================
\section{Introduction}

As never before in history, there are large volumes of data available today, for example, internet data, engineering signals or medical images. It is often difficult to extract valuable information from these datasets using traditional statistical methods. In addition, in most cases, we want to know the relationship between the components of the data, which is usually taken as a pairwise relationship. It is therefore natural to approach the problem from the point of view of graphs. However, in most real systems, the relationship between the components is not bipartite. In these cases, what is usually done is to compress these complex relationships between pairs, without considering that this generates a loss of valuable information. To overcome this problem, Berge developed the hypergraph theory \cite{berge1967graphes,berge1973graphs}, which represents the multiple component relationships in a system.
Hypergraph theory has been used in recent years in multiple applications from signal analysis \cite{zhang2019introducing,barbarossa2016introduction}, study of chemical reaction networks \cite{temkin2020chemical,konstantinova2001application} to biological networks \cite{feng2021hypergraph,klamt2009hypergraphs}. It has also been used in the study of clinical pathologies such as cardiac problems \cite{choudhury2019champs}, neurodegenerative diseases \cite{zhu2018dynamic} and epilepsy detection \cite{guo2021detecting}. Numerous applications of hypergraphs can be found in the area of machine learning, from image classification \cite{yu2012adaptive}, use of genetic algorithms \cite{raman2017efficient}, to detection of covid-19 in CT images. For a more in-depth review of hypergraph applications, see \cite{ouvrard2020hypergraphs}.

In this paper, we use hypergraph theory to study the metric structure of a data set. For this we use the idea of filtration coming from algebraic topology. This implies not having a fixed hypergraph for the data, but a set of hypergraphs which are formed based on a variable parameter. The idea is that dynamic hypergraph approach provides a powerful tool to infer robust qualitative and quantitative information about the structure of the data. This is done by taking as a basis a finite set of data --which can be points, signals, distributions or images, among others-- with a notion of distance or similarity between them. This distance can be induced by the metric in the ambient space (e.g. the Euclidean metric when the data are embedded in $\R^n$) or come as an intrinsic metric defined by a pairwise distance matrix. The definition of the metric on the data is usually guided by the application. Once we have chosen the metric space to work with, we define balls of radius $r$ centered at each data point we have. Subsequently, we increase the radius of the balls centered at any data point until we cover the entire set of the data. The data that fall into the balls are the ones that form the hyperedges of the hypergraphs. For each radius, we generate the associated hypergraph and apply a quantifier. This filtering method allows us to study the relationships between the data that remain unchanged as the radius changes. Different data topologies will present different quantifier values, allowing us to distinguish between different systems. 
In particular, in this study, we used five sets of points in the space  generated with different distributions and four sets of data coming from dynamical systems. The data were studied in five different metric spaces. In all cases, the filtering method was applied to generate the dynamic of the hypergraphs. The number of hyperedges was computed as a quantifier for each hypergraph as a function of the radius $r$ and the $L^1$-norm and the Sovolev discrete seminorm were calculated for each generated curve in order to collapse all the information into a single number. The results obtained show that it is possible to differentiate all the data sets analysed, both point distributions and dynamical systems. However, these results depend on the metric space used to generate the distance matrix of each system. This freedom in the choice of the underlying metric in the space containing the data set provides flexibility and robustness to the method. 

%======================================================================================================
\section{Graph, hypergraphs and covering.}
\label{Hyper_Theory}
\subsection{The general setting.}

Let $\chi$ be a set. We may  think of $\chi$  as a very large set containing all our possible data points. $\chi$  can  be the set of points in some euclidean space $\mathcal{R}^{n}$ or even some functional space when our ``data points'' are signals or images. In classical mathematical analysis such spaces can be taken to be Hilbert spaces, Banach spaces or even metric space. For the sake of generality, which will allow us to consider non symmetric ``metrics'' such as the Kullbach-Leibler divergence, we shall not even assume that $\chi$ is a metric space. Instead, we shall start considering a much more general metric-like structure an $\chi$.

\begin{defin}
A function $\delta:\chi~ \times~  \chi \longrightarrow \mathbb{R}^{+}_{0} $ shall called a ``protometric'' if $\delta$ satisfies $\delta(x,y)=0$ if and only if $x=y$. The $\delta$-ball centered at $x\in \chi$ with radious $r>0$ is defined by: $\mathcal{B}_{\delta}(x,y)= \left\lbrace\  y \in \chi :\delta(x,y)<r \right\rbrace $.
\end{defin}

Observe that with the above definition of protometric, in general $\mathcal{B}_{\delta}(x,r)$ does not coincide with the set $\left\lbrace\  y \in \chi :\delta(x,y)<r \right\rbrace $ since  $\delta$ may not be symmetric. Of course every metric in $\chi$ is a protometric. Recall that a metric $d$ in $\chi$ satisfies the additional properties:
\begin{enumerate}[a.]
    \item $d(x,y)=d(y,x)$ for every choice of $x$ and $y$ in $\chi$, and
    \item $d(x,z)\leq d(x,y)+d(y,z)$ for every choice of $x$, $y$ and $z$ in $\chi$.
\end{enumerate}

Let $\mathcal{V}=\left\lbrace x_{1},x_{2},...,x_{n}\right\rbrace$ be a finite sample of points in $(\chi,d)$. For any positive $r$ and every $i=1,2,...,n$; set
$$ e_{i}(r)=\mathcal{B}_{\delta}(x_{i},r) \cap \mathcal{V} =\left\lbrace x_{j}\in \mathcal{V} :\delta (x_{i},x_{j}) < r \right\rbrace, $$to denote the  hyperedge centered at $x_i$ with radius $r>0$.

Set $\mathcal{E}(r)$ to denote the family of all the hyperedges with radius $r>0$ fixed. Notice that the number of hyperedges in $\mathcal{E}(r)$ is some number $m(r)$ between one and $n$.

\begin{prop}
For each $r>0$, the couple $(\V,\E(r))$ is an \textit{hypergraph} with the additional property $x_{i}\in e_{i}(r)$ for every $i=1,2,...,n$.
\end{prop}

%\begin{proof}
%It is clear that each $e_{i}(r)$ is a subset of  $\V$. Moreover, since  $\delta(x_{i},x_{i})=0 <r$ we have that %$x_{i} \in e_{i}(r)$  for every i. $\diamondsuit$ 
%\end{proof}
%
If we consider this hypergraph structure as a function of $r>0$, we have some basic and elementary properties.

\begin{prop}
Let ($\V,\E(r)$) as before. Then

a) $e_{i}(r_{1})\subset e_{i}(r_{2})$ if $0<r_{1} \leq r_{2} < \infty$;

b) $e_{i}(r)= \left\lbrace\ x_{i}\right\rbrace$  if $0<r\leq \underset{j}{min}~\delta (x_{i},x_{j}) $;

c) $e_{i}(r)=\V$ if $r > \underset{j}{max}~ \delta(x_{i},x_{j}).$
\end{prop}
In other words, we have a family of hypergraphs starting at the trivial isolated point of $\V$  and finishing at the trivial full hypergraph whose only hyperedge is $\V$ itself.

%Notice also that for $r>0$ fixed it could happen that $e_{i}(r)=e_{j}(r)$ for some $i,j \in \left\lbrace\1,2,...,n\right\rbrace$. Even when they coincide as subsets of $\V $ we preserve the whole family of hyperedges as a sequence \textcolor{red}{Hugo en esta parte no se si poner la ultima expresión ya que mas adelante digo que si son iguales $e_i(r)=e_j(r)$ elimino una para formar la matriz de incidencia}
%$$\E(r)=(e_{1}(r),...,e_{n}(r))$$
%of the $n$ $e_{i}(r)$'s, because their coincidence is providing or recording some important geometric aspects of the points in $\V$. 

Now we can generate an adjacency  $n \times n$ matrix $\A(r)$ associated to the hypergraph $(\V,\E(r))$
which is given by
$$\A(r)=
\begin{pmatrix}
1 & \mathbb I_{e_{2}(r) } (x_{1}) &...& \mathbb I_{e_{n}(r)}(x_{1})\\
\mathbb I_{e_{1}(r)}(x_{2}) & 1 &...& \mathbb I_{e_{n}(r)}(x_{2})\\
\mathbb I_{e_{1}(r)}(x_{3}) & \mathbb I_{e_{2}(r)} (x_{3}) &...& \mathbb I_{e_{n}(r)}(x_{3})\\
.&.&.\\
.&.&.\\
.&.&.\\
\mathbb I_{e_{1}(r)}(x_{n}) & \mathbb I_{e_{2}(r)}(x_{n}) &...& 1\\
\end{pmatrix} =\left( \mathbb I_{e_{j}(r)} (x_{i})\right)_{\underset{j=1,...,n}{i=1,...,n}}$$
while $\I_{\theta}$ is the indicator functional of $\theta$, i.e  $\mathbb I _{\theta}(x)=1$ if $x\in \theta$  an $\I _{\theta}(x)=0$ is   $ x \notin \theta $. Notice that $(\chi ,d)$ is a metric space, then $\A(r)$ is symmetric for every $r>0$. In fact, since $d(x,y)=d(y,x)$, we readily have that 
$\mathbb I_{e_{j}(r)}(x_{i})=\mathbb I_{e_{i}(r)}(x_{j})$,
because $d(x_{i},x_{j})<r$ if and only if $d(x_{j},x_{i})<r$.

\begin{figure}[!htb]
\centering
\includegraphics[width =1\columnwidth]{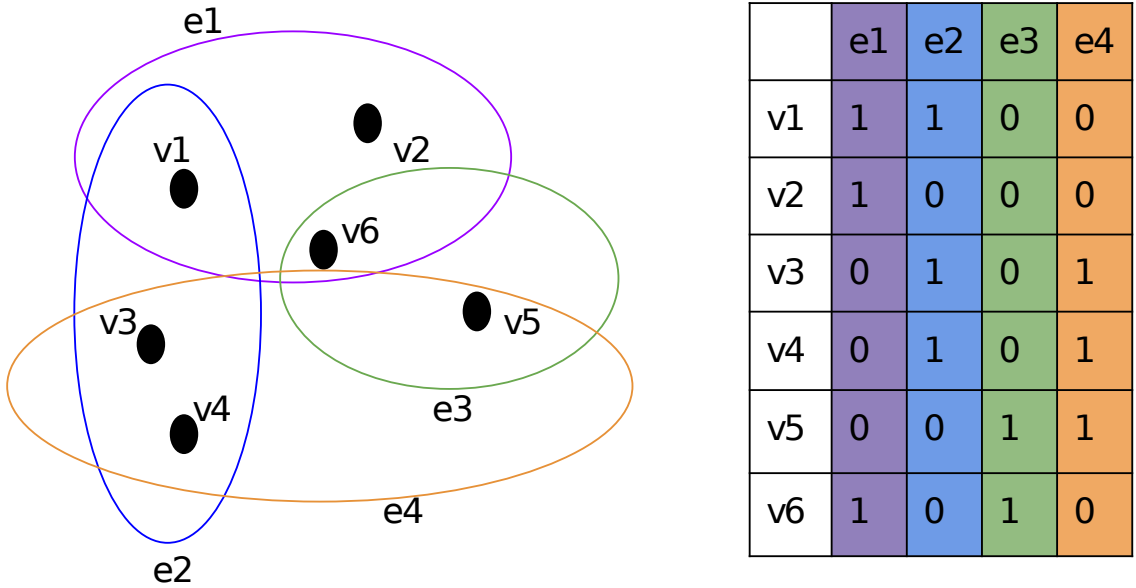}
\caption{ Incidence matrix for a hypergraph formed by 6 vertices and 4 hyperedges.}
\label{Fig_hyp_incidence}
\end{figure}

Notice also that for $r>0$ fixed it could happen that $e_{i}(r)=e_{j}(r)$ for some $i,j \in \left\lbrace 1,2,...,n\right\rbrace$. In this case the repeated column of the matrix (hyperedges) are eliminated, resulting in a new matrix $n \times m(r)$ with $m(r) \leq n$ call incidence matrix $\II$. Figure \ref{Fig_hyp_incidence} show an example of a hypergraph and it's incidence matrix associated.

\begin{defin}
Giving an incidence matrix $\II(r)$ $m(r) \times n$ associated to the hypergraph $(\V,\E(r))$ with $r>0$, we define the \textbf{degree of hyperedges } $\Delta^e(r)$ as  
\begin{equation}
    \Delta^e(r)=m(r)
    \label{eq:Nhyperedges}
\end{equation}
%
%and the \textbf{degree of vertex} $\Delta^v_{i}(r)$  for a edge $e_i$  as: 
%\begin{equation}
    %\Delta^v_{i}(r)=\sum _{j=1}^{m} \I_{e_{j} ^ {x_{i}}(r)}~~ for~~ i \in \left\lbrace 1,2,3,....,n\right\rbrace .
%\end{equation}
\end{defin}

This measure will be used to characterise the hypergraphs generated for the different values of $r$.

%==================================================
%                   Methods
%==================================================

\section{Methods}

\subsection{Data }
\label{sec:data}
\begin{figure}[!htb]
\centering
\includegraphics[width =1\columnwidth]{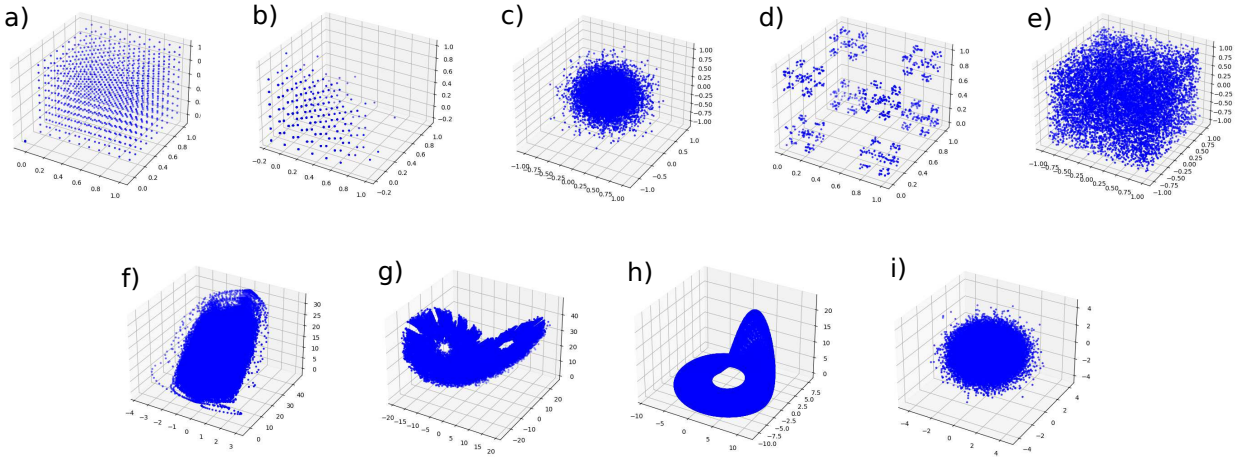}
\caption{Examples of different set of data used in the analysis. Points distribution / arrays (top)   a) Lattice, b) Poisson, c) Normal, d) Fractal and e) Uniform.  Dynamical systems (bottom) f) Complex Butterfly, g) Lorenz map, h) Rossler map, i) White noise. }
\label{Fig_DataPoint}
\end{figure}

In this paper, we used two types of data sets. The first one are sets of points in $ \R^3$ generated under different distributions: i) Normal, ii) Uniform, iii) Poisson and generated with a specific structure iv) Lattice and v) Fractal (see Figure \ref{Fig_DataPoint}(top)). For each set, we generated 100 realizations of $1000$ points each one.
The second type of data set belongs to 3-dimensional chaotic systems with different topologies: i) Rossler map, ii) Complex Butterfly map, iii) Lorenz map and a iv) white noise (see figure \ref{Fig_DataPoint} (bottom)). For each system, we generated $100$ sequences of $10000$ data each one, and then subsampled to 1000 data. All  datasets generated are described in depth in Appendix \ref{Apendix:Point_dist}.

\subsection{Basic Metrics}

As we explained in Section \ref{Hyper_Theory},  giving a  set of points $\chi$ we can use a metric  to measure the distance  between all pairwise  components of the set, giving a $n \times n$ distance matrix $\D_{\chi}$. 
In the sets considered in this paper we used five different metrics that we proceed to define explicitly.

\begin{defin}
Giving a set of point $\chi \in \R^n$ with $\chi= \{ \textbf{x}_1,\ldots,\textbf{x}_m \}$ we define the component $i,j=1,\ldots,m$ with $j>i$  of the matrix distance $\D_{\chi}$ as:\\
\begin{enumerate}
    \item Euclidean: $$ \D_{\chi}^{Ecu}(i,j)=\sqrt{\sum_{l=1}^{n}(\textbf{x}_i^l-\textbf{x}_j^l)^2},~~  where ~~x_i=(x^1_i,\ldots,x^n_i)$$

    \item Chebyshev: $$ \D_{\chi}^{Cheb}(i,j)=\underset{l=1,\ldots,n}{max} |\textbf{x}_i^l-\textbf{x}_j^l|$$

    \item Cityblock: $$ \D_{\chi}^{city}(i,j)=\sum_{l=1}^{n}|\textbf{x}_i^l-\textbf{x}_j^l|$$
    
     \item Minkowski: $$ \D_{\chi}^{Mink}(i,j)=\left ( \sum_{l=1}^{n}|\textbf{x}_i^l-\textbf{x}_j^l|^p \right )^\frac{1}{p}, ~~ 1\le p < \infty$$
    
    \item Parabolic: $$ \D_{\chi}^{Par}(i,j)= \underset{l=1,\ldots,n}{max}|x_i^l-x_j^l|^{\alpha_l}~~~with~~~0< \alpha_l \leq 1, ~~l=1,\ldots,n.$$

\end{enumerate}

\end{defin}

\subsection{Steps to built and characterise the hypergraphs $\H(r)$}
\label{sec:algorithm}

\begin{figure}[h]
\centering%
\includegraphics[width =0.8\columnwidth]{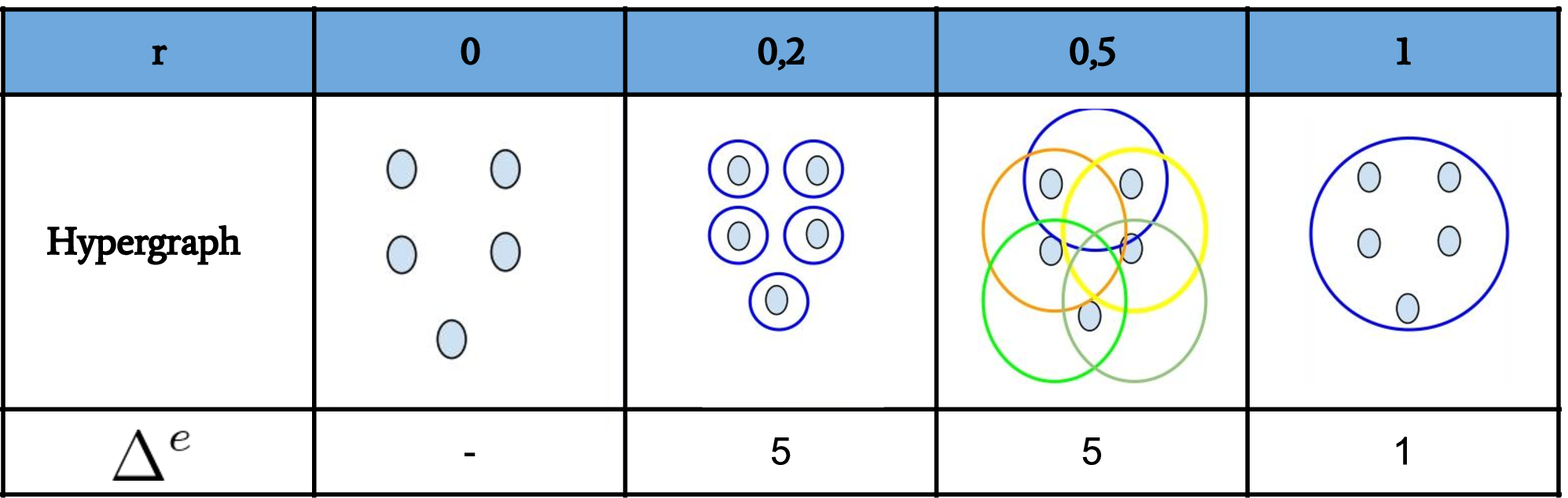}
\caption{Steps  to build and measure the feature $\Delta^e$ of the hypergraph for different values of $r$.}
\label{fig:algorithm}
\end{figure}

Let us now proceed to summarise the algorithm used to analyse a set of data based on the characterization of hypergraph. 
Given a data set $\chi= \{ x_1,...,x_m \}$ we generate and characterise the hypergraph associate to $\chi$ as follow,  (see Figure \ref{fig:algorithm}).

\begin{algorithm}
  \caption{Hypergraph filtration method }
  \begin{algorithmic}[1]
    
    \State Define the distance $d$ over the component of the set $\chi$. 
    \State Build the distance matrix $\D_{\chi}$.
    \State For each $x \in \chi$ define a ball centred in $x_i$  $\B(x_i,r)$ with  $0<r<1$. 
    \State Build the matrix $\M$ for each $r$ as 
    
    $$\M_{i,j}(r)= \left\{ \begin{array}{lcc}
             1~if~d^k(x_{i},x_{j}) \in \B(x_i,r) \\ 
             0~if~d^k(x_{i},x_{j})  \notin \B(x_i,r)
            
   \end{array}
   ~~for~~i,j =1,\ldots,m,~i<j. 
   \right.$$
    
    \State  Obtain the $n \times  m(r),~m(r) \leq n$ incidence matrix $\II(r)$  eliminating the repeated columns of the matrix $\M(r)$.
    \State Compute the degree of  hyperedges $\Delta^e(r)$ 
    
\end{algorithmic}
\end{algorithm}

%==================================================
%                   Result
%==================================================
\pagebreak

\section{Results}

\begin{figure}[h]
\centering
\includegraphics[width =1\columnwidth]{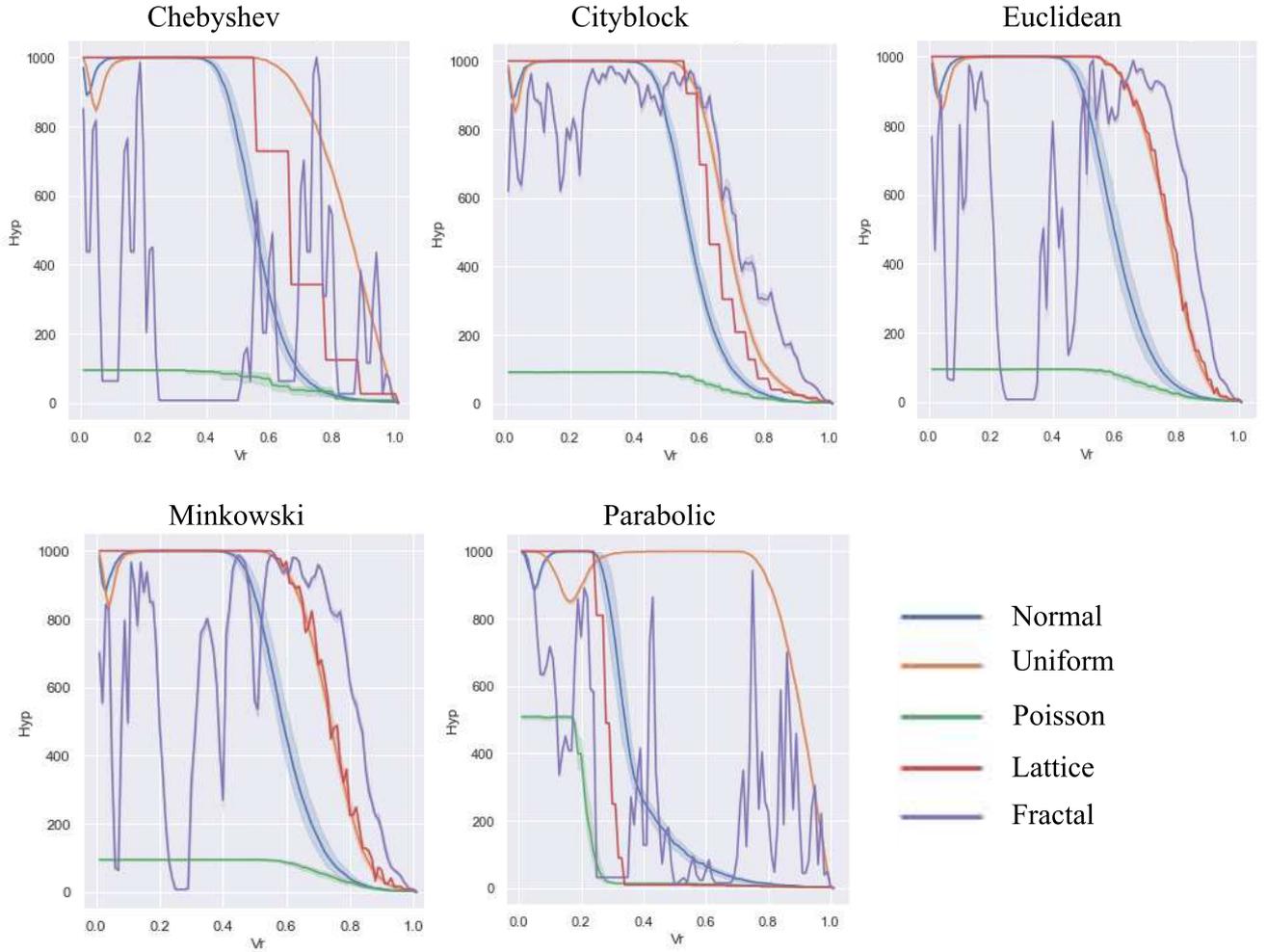}
\caption{Analysis of the number of hyperedges  versus parameter $r$, for $\R^3$ data points distribution. Each subplot represents the distance metric used to obtain the hypergraph.
}
\label{Fig_result_Dist_point}
\end{figure}

\begin{figure}[h]
\centering
\includegraphics[width =1\columnwidth]{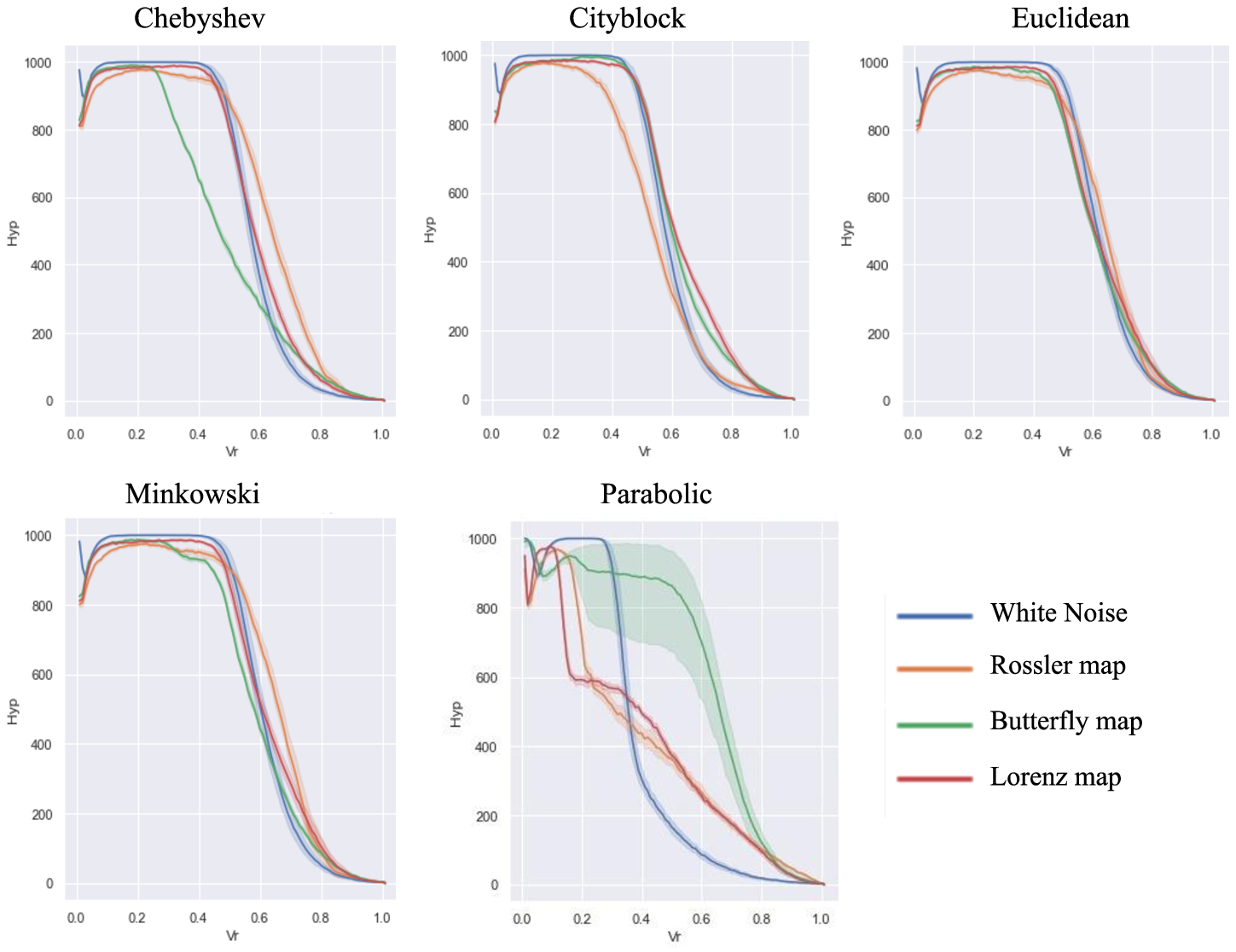}
\caption{ Analysis of the number of hyperedges versus parameter $r$, for three dynamical system (Lorenz map, Rossler map and Complex Butterfly map) and white noise. Each subplot represents the distance metric used to obtain the hypergraph.}
\label{Fig_result_Dyn_sys}
\end{figure}

We analysed the distributions of points and dynamical systems presented in Section \ref{sec:data} using the filtration algorithm propose in the Section \ref{sec:algorithm}.
In Figure \ref{Fig_result_Dist_point} and  \ref{Fig_result_Dyn_sys} we show  the Hypergraph degree $\Delta^e$ vs $r$ for distribution points and dynamical systems respectively. 
The straight line represent the mean value over $100$  realisation and the shadow band is the standard deviation.  Each subplot correspond a different distances used to obtain the hypergraph. 

In Figure \ref{Fig_result_Dist_point} it can be observed the results for the case of $\R^3$ data point distributions. We can see that the analysis of the number of hyperedges $\Delta^e$ for all the distance can clearly distinguish between the fives types of point distribution, being more remarkable the Chebyshev and Parabolic distance. 

%% cubica
Particularly it is seen that for the isotropic metrics ($\D^{Euc},~\D^{Cheb},~\D^{City},~\D^{Mink}$), the lattice array  starts with the number of hyperedges similar to the number of vertices, and remains constant until $r$ approaches $r=0.58$, where the number of hyperedges decreases until it reaches zero. Depending on the distance used, this decrease is not uniform, being fluctuating for the Euclidean and Minkowski, and stepwise for the Chebyshev and cityblock distance. On the other hand, for the parabolic distance, the behaviour is very different,  starting with very low number of hiperedges $\Delta^e \sim 150$ and decaying to zero in ($r \sim 0.28$).

In the case of the Normal distribution, the number of hyperedges is similar to the number of vertices for small $r$. For $r \sim 0.06$ a drop in $\Delta^e$ is observed but it quickly recovers the initial values. Finally, there is a uniform drop in the number of hyper edges reaching zero for $r=1$, this occurs for $r \sim 0.6$ in the symmetric distance and $r \sim 0.3$ for the parabolic distance.

Uniform distribution present a similar behaviour as the Normal distribution. However the uniform decrease in the number of hyperedges start before in $r \sim 0.6$ for all the metrics.  

The Poisson distribution showing a higher data clusterization giving a lower number of hyperedges even for initial values of $r$. These values remain constant until $r \sim 0.06$ where decrease uniformly to zero. This behaviour occurs for all the metrics.  

In the case of fractal distribution, the number of hyperedges depends strongly in the $r$ values, and the behaviour of the curves is very different for the distinct metrics. Particularly, the Chebyshev and Parabolic distances showing constant periods with very low numbers of hyperedges alternating with peaks with high numbers $\Delta^e$. Similar fluctuations are presented in the other metrics in the range between $0 < r \le 0.7$, before $r \sim 0.7$ the $\Delta^e$ decay to zero, similar to the other distribution. 

%---------------------- Figure Data Distribution --------------
%-------------------------------------------------------

In the second instance, we analysed the dynamical systems presented in the Section \ref{sec:data}. Figure \ref{Fig_result_Dyn_sys}  shows the analysis of the number of hyperedges  vs $r$ parameter for the dynamics system. For the $\Delta^e$ study, the Chebishev and Parabolic distance (with $\alpha^1=1,~\alpha^2=1/2,~\alpha^3=1/2 $) are the best at differentiating the four dynamical system. On the other hand, the Euclidean distance could not see any differences between them. For the fives metrics, white noise starts with the similar number of hyperedges than vertices for small $r$ and shows a drop in $r\sim0.06$ but fast returns to original values. Then, remains with higher values of $\Delta^e$ until $r\sim0.5$ for isotropic  metrics and $r\sim0.3$ for parabolic distances when decrease uniformly to zero in $r=1$. For isotropic metrics Rossler map begin  with $\Delta^e <N_{vert}$ but promptly increases near to the maximum at $r\sim0.2$ and remaining until $r\sim0.6$ then decay to zero. However, for the parabolic distance the behaviour of the curve is very difference decreasing very sharp at $r\sim0.15$ and changing the slope at $r\sim0.2$. Complex butterfly and Lorenz maps present a similar behaviour for the isotropic metrics except for Chebyshev where the slope changes a little between then. But for Parabolic distance the differences between this two chaotic maps are very remarkable. Similar situation we have for Lorenz map where the only distance that can clearly distinguish for the others dynamical systems is the parabolic distance. 

%---------------------- Figure Data Distribution --------------
%-----------------------------------

\subsection{Distance quantification}

In the previous section, we saw that distinct distribution or dynamics could be differentiated on the base of the shapes of the curves traced by the function $\Delta^e(r)$. In this section, we  quantified more accurately the result obtained with $\Delta^e(r)$. For that aim we used two metric, $L^1$-norm  $(\L)$ \cite{gradshteyn2014table} and the discrete Sobolev semi-norm of order $1$ $(\S)$. 
%Been $\textbf{v}=\{v(r_i),...,v(r_m) \}$ the vector of values correspond with $\Delta^e(r)$ 
The $L^1$-norm is defined as 
$$ \L=\sum_{i=1}^m |\Delta^e(r_i)| $$
and the Sobolev seminorm as:
$$ \S=\sum_{i=2}^{m}  \frac{|\Delta^e(r_i)-\Delta^e(r_{i-1})|}{r_i-r_{i-1}} .$$
The idea behind the use of these two metrics is the possibility to quantify the content of the curves by its area ($\L$) and the changes in their slopes ($\S$) 

Moreover, to measure a significant distance between two different systems we define the following pipeline. Having a group of curves  $\V=\{ \Delta^e(r)_1,...,\Delta^e(r)_n \}$ and $\hat{\V}=\{ \hat{\Delta^e(r)}_1,...,\hat{\Delta^e(r)}_n \}$ corresponding with $n$ realization of each system (for example $n$ different white noises), we estimate the mean values $\mu(r)$ and standard deviation $\sigma(r)$ for each group. Then, we calculate the distance $d^{\V,\hat{\V}}(r)$ between the two group as follows:\\
If $\sigma(r) < \hat{\sigma}(r)$ then
$$
 d^{\V,\hat{\V}}(r) =
  \begin{cases}
     ( \mu(r)-\sigma(r)) - ( \hat{\mu}(r)+\hat{\sigma}(r))    & \quad~ \text{if } (\mu(r)-\sigma(r)) > (\hat{\mu}(r)+\hat{\sigma}(r))\\
    0  & \quad \text{if } ~ (\mu(r)-\sigma(r)) \leq (\hat{\mu}(r)+\hat{\sigma}(r))
  \end{cases}
$$
or $\sigma(r) > \hat{\sigma}(r)$ then
$$
 d^{\V,\hat{\V}}(r) =
  \begin{cases}
     (\hat{\mu}(r)-\hat{\sigma}(r)) - ( \mu(r)+\sigma(r))    & \quad~ \text{if } (\hat{\mu}(r)+\hat{\sigma}(r)) > (\mu(r)-\sigma(r))\\
    0  & \quad \text{if } ~ (\hat{\mu}(r)-\hat{\sigma}(r)) \leq (\mu(r)+\sigma(r))
  \end{cases}
$$
Finally we quantify the difference applying the two norms described before $L^1$-norm and Sovolev semi-norm of order $1$ to the distance vector $d^{\V,\hat{\V}}(r)$.

Table \ref{table:PointDist_L_S} shows the $\L$ and $\S$ mean and standard deviation values over $\Delta^e$ curves for the different points distribution. As we could saw in the figure \ref{Fig_result_Dist_point}, the Parabolic and Chebichev distance are the best metrics to differentiating the fives distribution -this is reflected in the wide distribution of $\L$ and $\S$ values in the table-. Cityblock and Euclidean distances could not  discriminate well the five cases, and Minkowisky only can discern the distribution just for $\S$ metric. For all the distances we can remark the high difference between fractal $\S$ values  array with the others distribution, this behaviour is not reflected in $\L$ values. This facts reflects the fractal intrinsic character of the fractal.   
In table \ref{table:PointDist_Par_L} and \ref{table:PointDist_Par_S} we measure the distance $\L$ and $\S$ respectively between the distribution for the Parabolic distance (the other distances are shown in supplementary material). For $\L$ values, Normal distribution  and lattice are the most similar, and Poisson and uniform present the highest differences. In $\S$ distance between Poisson  and  lattice  become more similar, and the lattice and fractal have the highest distance .

For the dynamical systems, the $\L$ and $\S$ over $\Delta^e$ is presented in the table \ref{table:DynSys_L_S}. In this case, the differentiation between systems is less clear than before. Chebyshev is the ones which shown significant difference $\L$ and parabolic in $\S$ values. Measuring the distances between the systems we can better appreciate their differences. For that we measure  $\L$ para el caso de la Chebyshev (table  \ref{table:DynSyst_Cheb_L}) and the  $\S$ for the parabolic case (\ref{table:DynSyst_Par_S}).

%%%%%%%%%% --------------------TABLE 1 point distribution L and S values --------------------------------------------------------------    
\begin{table}[]
    \centering
    \input{tables/Tabla1}
    \caption{Mean values and standard deviation for the $L^1$-Norm $(\L)$ and Sobolev seminorm $(\S)$ applied over the $\Delta^e(r)$ curves belong to the different 3D point distribution.}
    \label{table:PointDist_L_S}
\end{table}

%%%%%%%%%% --------------------TABLE 2 Dynamical System L and S values --------------------------------------------------------------

\begin{table}[]
    \centering
    \input{tables/Tabla2}
    \caption{Mean values and standard deviation for the $L^1$-Norm $(\L)$ and Sobolev seminorm $(\S)$ applied over the $\Delta^e(r)$ belong to the different dynamical system.}
    \label{table:DynSys_L_S}
\end{table}
%%%------------------------ Table 3 Parabolic Distance L between point distribution ---------------------

\begin{table}[]
    \centering
    \input{tables/Tabla3}
    \caption{$\L$ distance matrix for parabolic metric between 3D points distribution. }
    \label{table:PointDist_Par_L}
\end{table}

%%%------------------------ Table 4 Parabolic Distance S between point distribution ---------------------

\begin{table}[]
    \centering
    \input{tables/Tabla4}
    \caption{$\S$ distance matrix  for parabolic metric between 3D points distribution.}
    \label{table:PointDist_Par_S}
\end{table}

%%%------------------------ Table 5 Parabolic Distance L between point distribution ---------------------

\begin{table}
    \centering
    \input{tables/Tabla5}
    \caption{$\L$ distance matrix for Chebyshev  metric between dynamical system.}
    \label{table:DynSyst_Cheb_L}
\end{table} 
%
%%%------------------------ Table 6 Parabolic Distance S between point distribution ---------------------
\begin{table}
    \centering
    \input{tables/Tabla6}
    \caption{$\S$ distance matrix for parabolic distance between dynamical system.}
    \label{table:DynSyst_Par_S}
\end{table}

%==================================================
%                   Discussion 
%==================================================
\section{Discussion}  

In this works we introduce a novel approach to study different  datasets models  using quantifiers obtained by  the filtration  evolution of the  hypergraph. Particularly, we study the cases of five distributions of  points and four dynamical systems.  Our method shows the strength to  extract information about the metric of the dataset,  being able to distinguish and quantify all the distribution and the dynamical systems studied. 

The study of the $\Delta^e(r)$ curves allow extract quantify in simple way the information extracted for the data. High values of the $\Delta^e(r)$ represent a low data clustering. When the radius $r$ are small, the number of adjacent points tent to be zero, for example in cases as Normal or Uniform distribution $\Delta^e(r)=N_{data}$ --the ball of each point only contain itself. However, in other cases as Poisson, fractal distribution or Rossler map, for the initial radius $r=0.01$ the $\Delta^e<<N_{data}$, showing that the data are more clustered.

In all the example, with exception of fractal data, we can see that the initial values of $\Delta^e(r)$ remains constant as the $r$ increase, drooping to zero at some critical $r_c$ which depends of the distribution or dynamical system; the decrease in $\Delta^e(r)$ is maintained as $r$ increases until its reach the value $r=1$ --where  a single hyperedge  contain all points. These behaviour is expected because as the $r$ increase, the number  hyperedges containing the same points increase --remember when two hyperedges have the same points we consider only one. However, depending of the distance used between points, each database present different curve shapes. For example, the lattice distribution present a staircase decay for the Chebichev and Cityblock distance.  The fractal behaviour  is very different from the others distribution due to why present  picks and  valleys as the value of $r$ grows, this occurs due to the fractal nature of the distribution of the points where both clusters of points and large empty spaces exist in space.  
The difference between point distribution and dynamical systems depend in the metric used to generate the hypergraph. In both cases the Parabolic distance exhibit better results extracting information from the hypergraph topology which can be used to quantify and discern between then. 
The $L^1$-Norm $(\L)$ and Sobolev seminorm $(\S)$ are presented as useful metrics for quantifying the curves $\Delta^e(r)$ obtained filtration method. Moreover, the pipeline introduced to measure a distance  between two curves allows to unveil differences between the curves which cannot be obtained by comparing only the intrinsic values (both $\L $ and $\S $) of each curve. 

In this work we use  the number of hyperedges as a quantifier of the hypergraph, however this not means that no exist other quantifier which gives different or complementary information  about the hypergraph, for example obtain information about the vertices in the hyperedges. In the future works, we going  to implement others quantifiers based in different attribution of the hypergraph. 

Although in this work we have focused on the analysis of datasets of points, as described in section \ref{Hyper_Theory}, this method applies to any dataset within a defined metric space such as signals, images, distributions, etc. In future work, we propose to extend this pipeline to the analysis of different data sets obtained from real life.

%============================================================
%	REFERENCE LIST
%-============================================================

\bibliography{bibliography.bib}
\bibliographystyle{unsrt} 

\appendix

\section{Point Distribution Generation}
\label{Apendix:Point_dist}
\subsubsection*{Normal distribution}
A discrete random variable $x$ is said to have a Normal distribution if:
$$p(x,\mu,\sigma^{2})=\dfrac{1}{\sigma \sqrt{2 \pi }} e^{\dfrac{-1}{2} .\left( \dfrac{x-\mu}{\sigma }\right)^{2}}$$
where  $\mu$ is the location parameter, and it is going to be equal to the arithmetic mean and is the location parameter, and it is going to be equal to the arithmetic mean and $\sigma ^{2}$ is the standard deviation. In our work for the different realisation we use the function  {\ttfamily random.normal} in  numpy Python  package.

\subsubsection*{Poisson distribution}
A discrete random variable $x$ is said to have a Poisson distribution if:
$$p (x = k) = \lambda^{k}. \frac{e ^{- \lambda} }{k!} $$
Where $k$ is an integer ($ k \geq 0 $) and $\lambda$ is a positive real number. The Poisson distribution describes the probability of encountering exactly $k$ events in a time span if the events occur independently at a constant rate $\lambda$. To generate the distribution we use the function {\ttfamily random.poisson} in  numpy Python  package. 

\subsubsection*{Uniform distribution}
The uniform  Distribution is given by the formula:
$$p(x,a,b)= \left\{  \begin{array}{lcc}
             \dfrac{1}{b-a} ~~ for~~ a \leq x \leq b\\
             0 ~~~~~~~ for ~~ x < a~ or~ x > b\\
             \end{array}
   \right.$$

In our work we use $a=-1$ and $b=1$ and user the function {\ttfamily random.uniform} in  numpy Python  package. 

\subsubsection*{Lattice array}
A lattice is an ordered array of points describing the arrangement of particles that form a crystal. In our work we build the lattice defining a 3-dimensional cube and putting the data in a \textbf{x}-distance position. 

\subsubsection*{Fractal array}

For the fractal array we generate $N=1000$ 3D points. The coordinates $\mathbf{X}=[x^1,x^2,x^3]$ of the each point  follows the next pipeline. 
For each coordinate $x^l$ we generate a vector of length $L=100$,  $\mathbf{x}^l=\{x^l_1,\ldots,x^l_{100} \}$ with $x^l_i$ randomly chosen between $\{0,3\}$ --for example $\mathbf{x}^1=\{0,3,0,0,0,3,0,...,0\}$--, then we measure the $x^l$ values as: 

$$ x^l=\sum^L_{i=1} \frac{x^l_i}{4^i}~~~~ for~~l=1,2,3$$

\section{Dynamical System Generation}
\label{Apendix:Dyn_Sys}

\subsubsection*{Lorenz attractor}
The Lorenz attractor is defined as the dynamical system governed by the following system of equations:
$$ \left\{ \begin{array}{lcc}
             \dfrac{dx}{dt} = \sigma(y-x) \\
             \dfrac{dy}{dt} = -xz+rx-y \\
             \dfrac{dz}{dt} = xy -bz\ 
             \end{array}
   \right.$$
where we take the usual values  $\sigma=10$, $r=284$, $b=8/3$ of the parameters.
With the following initial conditions $x_{0}=0$, $y_{0}=-0.01$ and $z_{0}=9$

\subsubsection*{Rössler attractor}
The Rössler attractor is the attractor of the Rössler system, a system of three nonlinear ordinary differential equations

$$ \left\{ \begin{array}{lcc}
             \dfrac{dx}{dt} = -y-z \\
             \dfrac{dy}{dt} = x+ay \\
             \dfrac{dz}{dt} = b+z(x-c)\ 
             \end{array}
   \right.$$
where we take the usual values  of the parameters $a=b=0.2$, $c=5.7$.
With the following initial conditions  $x_{0}=-9$, $y_{0}=0$ and $z_{0}=0$. 

\subsubsection*{Complex butterfly attractor}
The Complex butterfly  attractor is a system of three nonlinear ordinary differential equations:
$$ \left\{ \begin{array}{lcc}
             \dfrac{dx}{dt} = a(y-z) \\
             \dfrac{dy}{dt} = -z~sgn(x) \\
             \dfrac{dz}{dt} =|x| -1\ 
             \end{array}
   \right.$$
   
where we take the usual values  of the parameters $a=0.55$, $c=5.7$.
With the following initial conditions  $x_{0}=0.2$, $y_{0}=0$ and $z_{0}=0$.

\end{document}

%% file: tables/Tabla1.tex
\scriptsize{
\begin{tabular}{llll}
\hline
System & Distance  & $\L$ ($\mu \pm \sigma$) & $\S$ ($\mu \pm \sigma$) \\
\hline
Lattice      &            & 68492$\pm$0      & 999$\pm$0       \\
Fractal    & Chebyshev  & 24364$\pm$125    & 10916$\pm$76    \\
Normal       &            & 56053$\pm$3597   & 1209$\pm$17     \\
Poisson      &            & 6579$\pm$539     & 96$\pm$4        \\
Uniform      &            & 82944$\pm$329    & 1321$\pm$16     \\
\hline
Lattice      &            & 65774$\pm$0      & 999$\pm$0       \\
Fractal    & Cityblock  & 66677$\pm$299    & 4835$\pm$205    \\
Normal       &            & 58774$\pm$2463   & 1211$\pm$21     \\
Poisson      &            & 6558$\pm$285     & 93$\pm$4        \\
Uniform      &            & 69315$\pm$1205   & 1299$\pm$21     \\
\hline
Lattice      &            & 74190$\pm$0      & 1527$\pm$0      \\
Fractal    & Euclidean   & 59941$\pm$301    & 8612$\pm$354    \\
Normal       &            & 60763$\pm$2970   & 1229$\pm$18     \\
Poisson      &            & 6584$\pm$648     & 109$\pm$6       \\
Uniform      &            & 72448$\pm$1240   & 1341 $\pm$24    \\
\hline
Lattice      &            & 76850$\pm$0      & 1143$\pm$0      \\
Fractal    & Minkowsky & 53260$\pm$361    & 9704$\pm$288    \\
Normal       &            & 60521$\pm$2541   & 1241$\pm$23     \\
Poisson      &            & 6744$\pm$ 767    & 142$\pm$27      \\
Uniform      &            & 60521$\pm$2541   & 1241$\pm$23     \\
\hline
Lattice      &            & 285780 $\pm$0   & 999$\pm$0     \\
Fractal    & Parabolic  & 29806$\pm$170  & 11558$\pm$76      \\
Normal       &            & 39081$\pm$2689 & 1261$\pm$27      \\
Uniform      &            &  87942$\pm$202  & 1335$\pm$21  \\
Poisson      &            & 11729$\pm$ 736  & 512$\pm$ 11   \\
\hline
\end{tabular}
}

%% file: tables/Tabla2.tex
\scriptsize{
\begin{tabular}{llll}
\hline
System & Distance  & $\L$ ($\mu \pm \sigma$) & $\S$ ($\mu \pm \sigma$) \\
\hline
Complex Butterfly  &  & 49037$\pm$ 626      & 1279$\pm$32     \\
Lorenz map            & Chebyshev  & 58370    $\pm$ 1042 & 1270$\pm$21     \\
Rossler map           &   & 61935 $\pm$1586     & 1322$\pm$40     \\
White Noise           &   & 57755$\pm$ 2228     & 1213$\pm$16     \\
\hline
Complex Butterfly  &  & 60686$\pm$1056      & 1268$\pm$25     \\
Lorenz map            & Cityblock  & 61540$\pm$1044      & 1284$\pm$17     \\
Rossler map           &  & 52628$\pm$1465      & 1260$\pm$16     \\
White Noise           &   & 58183$\pm$2490      & 1217$\pm$15     \\
\hline
Complex Butterfly  &    & 60133$\pm$818       & 1295$\pm$38     \\
Lorenz map            & Euclidean   & 63138$\pm$1582      & 1285$\pm$13     \\
Rossler map           &   & 61426$\pm$1548      & 1323$\pm$48     \\
White Noise           &   & 62095$\pm$1920      & 1247$\pm$14     \\
\hline
Complex Butterfly  &  & 57854$\pm$668       & 1298$\pm$38     \\
Lorenz map            & Minkowsky & 61012$\pm$1578      & 1290$\pm$24     \\
Rossler map           &  & 62710$\pm$1909      & 1317$\pm$29     \\
White Noise           &  & 60807$\pm$2289      & 1238$\pm$23     \\
\hline
Complex Butterfly  & & 61573$\pm$  13041   & 1340$\pm$ 29    \\
Lorenz map            & Parabolic  & 39240$\pm$ 992      & 1634$\pm$ 62    \\
Rossler map           &   & 39834$\pm$ 1973     & 1501$\pm$ 83    \\
White Noise           &   & 39342 $\pm$  2151   & 1265$\pm$  15  \\
\hline
\end{tabular}
}

%% file: tables/Tabla3.tex
\scriptsize{
\begin{tabular}{|l|l|l|l|l|l|}
\hline
          & Lattice & Fractal   & Norma       & Poisson     & Uniform     \\
          \hline
Lattice   & 0       & 23133 & 8460 & 16051 & 61770 \\
\hline
Fractal & 23133       & 0           & 23974 & 17676 & 57333 \\
\hline
Normal    & 8460       & 23974          & 0           & 23243 & 48398 \\
\hline
Poisson   & 16051       & 17676          & 23243           & 0           & 74607 \\
\hline
Uniform   & 61770      & 57333          & 48398           & 74607          & 0  \\
\hline
\end{tabular}
}

%% file: tables/Tabla4.tex
\scriptsize{
\begin{tabular}{|l|l|l|l|l|l|}
\hline
          & Lattice & Fractal  & Normal       & Poisson     & Uniform     \\
          \hline
Lattice   & 0       & 11928 & 1597 & 1487 & 2225 \\
\hline
Fractal & 11928       & 0          & 11195  & 11134 & 11414  \\
\hline
Normal    & 1597      & 11195         & 0           & 1639 & 2384 
\\
\hline
Poisson   & 1487       & 11134          & 1639           & 0           & 1857 \\
\hline
Uniform   & 2225       & 11414          & 2384          & 1857          & 0 \\
\hline
\end{tabular}
}

%% file: tables/Tabla5.tex
\scriptsize{

\begin{tabular}{|l|l|l|l|l|}
\hline
                  & White Noise & Complex Butterfly & Lorenz  & Rossler  \\
                  \hline
White Noise       & 0           & 4989              & 797    & 4137    \\
\hline
Complex Butterfly & 4989        & 0                 & 6975   & 10653   \\
\hline
Lorenz            & 797         & 6975              & 0      & 2468    \\
\hline
Rossler           & 4137        & 10653             & 2468   & 0 \\
\hline
\end{tabular}

}

%% file: tables/Tabla6.tex
\scriptsize{
\begin{tabular}{|l|l|l|l|l|}
\hline
                  & White Noise & Complex Butterfly     & Lorenz     & Rossler \\
                  \hline
White Noise & 0                 & 967 & 1603 & 1498 \\
\hline
Complex Butterfly & 976          & 0   & 1405 &925 \\
\hline
Lorenz           & 1603                 & 1405           & 0           & 784 \\
\hline
Rossler      & 1498                & 925          & 784          & 0 \\    \hline    
\end{tabular}

}